\documentclass[12pt]{article}
\usepackage{graphicx,color}
\usepackage{hyperref}
\usepackage{amssymb,amsfonts,amsmath,cancel,cite,multirow}
\usepackage[capitalise]{cleveref}

\newcommand{\pdiff}[3]{
\if 1#1   \frac{\partial #2 }{\partial #3 }
\else  \frac{\partial^#1#2 }{\partial #3^#1 } \fi}

\newcommand{\Tr}[0]{\mathrm{Tr}}
\newcommand{\tr}[0]{\mathrm{tr}}
\newcommand{\TIa}[0]{T^{(\alpha)}_I}
\newcommand{\TJb}[0]{T^{(\beta)}_J}
\newcommand{\CNcomp}[0]{C_N^{\mathrm{comp.}}}
\newcommand{\CMcomp}[0]{C_M^{\mathrm{comp.}}}
\newcommand{\qcomp}[0]{q_{\mathrm{comp.}}}
\newcommand{\Kcal}[0]{\mathcal{K}}
\newcommand{\Pcal}[0]{\mathcal{P}}
\newcommand{\Ocal}[0]{\mathcal{O}}
\newcommand{\lam}{\lambda}
\newcommand{\ep}{\epsilon}
\newcommand{\del}{\delta}
\newcommand{\Del}{\Delta}
\newcommand{\ovl}{\overline}
\newcommand{\eqs}[1]{\begin{equation}\begin{split} #1 \end{split}\end{equation}}

\begin{document}

\begin{titlepage}

\begin{flushright}
IPMU17-0043
\end{flushright}

\vskip 1.35cm
\begin{center}

{\large
\textbf{
Two-loop Anomalous Dimensions for Four-Fermi Operators in Supersymmetric Theories
}}
\vskip 1.2cm

Junji Hisano$^{a,b,c}$,
Takumi Kuwahara$^b$,
Yuji Omura$^a$,
and Takeki Sato$^b$\\

\vskip 0.4cm

\textit{$^a$Kobayashi-Maskawa Institute for the Origin of Particles and the Universe (KMI),
Nagoya University, Nagoya 464-8602, Japan}\\
\textit{$^b$Department of Physics,
Nagoya University, Nagoya 464-8602, Japan}\\
\textit{$^c$
Kavli IPMU (WPI), UTIAS, University of Tokyo, Kashiwa, Chiba
 277-8584, Japan}
\date{\today}

\vskip 1.5cm

\begin{abstract}
We derive two-loop anomalous dimensions for four-Fermi operators in supersymmetric theories using the effective K\"ahler potential.
We introduce the general forms in generic gauge theories and apply our results to the flavor-changing operators in (minimal) supersymmetric standard models.

\end{abstract}

\end{center}
\end{titlepage}

\section{Introduction \label{sec:intro}}

Searches for rare processes are one of the promising probes to test models for particle physics.
There are a lot of attempts to extend the Standard Model (SM) in order to explain the dark matter candidate, the origin of fermion masses and mixing, and so on.
Such models beyond the SM (BSMs) generally predict forbidden or suppressed observables in flavor physics, including proton decay and flavor-violating observables.
It is very important to find and estimate the deviations of those observables from the SM predictions in BSMs.
In the SM, the flavor-violating processes are predicted to be very small and are consistent with the experimental results, although there are still large uncertainties of the predictions in some processes.
Then, the contributions of the new physics should be also small, and the high accuracy of the predictions is required to prove the new physics.
Many experiments have been performed to survey new physics in flavor physics and succeeded in constraining extensions of the SM.
Besides, since there are many plans of next-generation experiments, it is required to prepare the theoretical predictions of the observables with high accuracy in the BSMs as well as the SM.

As one of the promising candidates for the BSMs, supersymmetry (SUSY) has been widely discussed to explain, for instance, the dark matter and the origin of the electroweak symmetry breaking.
In the supersymmetric grand unified theories proton decay is induced by new gauge bosons called as $X$ bosons.
In the minimal supersymmetric standard model (MSSM), flavor violation arises from soft supersymmetry breaking terms, and the flavor-violating operators are loop-suppressed as far as the $R$-parity is conserved (for instance, see Ref.~\cite{Gabbiani:1996hi}).
There are supersymmetric models that induce tree-level flavor changing neutral currents (FCNCs).
For instance, a $U(1)^\prime$-extended supersymmetric SM (SSM) predicts an additional gauge supermultiplet ($Z^\prime$), and the $Z^\prime$ interaction generates tree-level FCNCs if $Z^\prime$ has flavor non-universal couplings \cite{Langacker:2000ju,Demir:2005ti}.
The $U(1)^\prime$ extensions of the MSSM are expected in the context of string theory \cite{Cvetic:1995rj} and grand unified theories \cite{London:1986dk,Hewett:1988xc}.
It is also expected that the tree-level FCNC via $Z^\prime$ arises from the matter mixing in the $SO(10)$ GUT \cite{Hisano:2016afc,Hisano:2015pma}.

At low-energy scale, the rare processes are described by the effective operators that are generated by integrating out heavy fields.
Then, there is a large hierarchy between the energy scales;
the heavy mass scale and the scale where the matrix element is evaluated.
For instance, proton decay operators are generated at the grand unification scale ($\sim 10^{16}$~GeV), and the hadron matrix elements are estimated at the hadronic scale ($\sim 1$~GeV).
Besides, near the hadronic scale, the large QCD correction is expected due to the large coupling.
Therefore, renormalization group equation (RGE) analysis is necessary to predict such processes precisely.

Next-leading order (NLO) corrections for effective operators have been evaluated in some literature; \textit{e.g.},
the two-loop QCD anomalous dimensions for the flavor-changing four-Fermi interactions in the SM and the BSM are calculated in Ref.~\cite{Buras:2000if}, and the two-loop beta function for the effective operator which describes neutrino mass at low energy is given in Ref.~\cite{Antusch:2002ek}.
In the extended MSSM, the additional gauge symmetry is possibly broken above the SUSY breaking scale, where the superpartners of the SM particles reside.
Global and/or gauge symmetry breaking at an intermediate scale has been discussed to solve the $\mu$-problem in the MSSM, the thermal inflation, and the neutrino masses; for instance, gauged $U(1)^\prime$ breaking \cite{Cleaver:1997nj,Erler:2002pr,Morrissey:2006xn} and Peccei-Quinn symmetry and/or $R$-symmetry breaking \cite{Murayama:1992dj,Kim:1994eu} are introduced.
If the tree-level FCNC arises from integrating out the massive vector superfields, it is important to include RGE effects between the masses of the massive vector superfields and the SUSY breaking scale.
The theories are supersymmetric between the scales, and thus it is valid to use the anomalous dimensions in the SUSY calculation.

In the SUSY theories, there can be no divergence in some quantum corrections, which is known as the non-renormalization theorem \cite{Seiberg:1993vc}.
In the $\mathcal{N}=1$ theories on the four-dimensional spacetime, interactions induced by the superpotential do not suffer from any quantum corrections except anomalous dimensions for constituent fields.
On the other hand, the K\"ahler potential is affected by other quantum corrections as well.
The two-loop effective K\"ahler potential is derived in a previous study \cite{Nibbelink:2005wc}.

It is shown in Ref.~\cite{Hisano:2013ege} that, for higher-dimensional K\"ahler potential terms, one-loop anomalous dimensions via gauge interactions are given with the group theoretical factors.
On the other hand, the two-loop anomalous dimensions have more complicated structure.
In our study, we show details of the anomalous dimensions for any K\"ahler-type interactions using the effective K\"ahler potential.
We find simple formulae for the anomalous dimensions of dimension-six interactions at two-loop level.

This paper is organized as follows:
in the next section, we briefly introduce the effective K\"ahler potential at two-loop level which is derived in Ref.~\cite{Nibbelink:2005wc}.
In \cref{sec:anomalous}, we derive the two-loop anomalous dimensions for the higher-dimensional K\"ahler potential terms.
We especially focus on the four-Fermi type operators.
We will show application of our results to $\Del F = 2$ operators and proton decay operators, which are generated by massive vector supermultiplets, in \cref{sec:application}.
Finally, we summarize our study in \cref{sec:conclusion}.

\section{Effective K\"ahler Potential \label{sec:Kahler}}

We briefly summarize the effective K\"ahler potential derived in Ref.~\cite{Nibbelink:2005wc}.
We now consider a generic $\mathcal{N}=1$ supersymmetric model on four-dimensional spacetime.
The model is constructed by (anti-)chiral superfields $\Phi$ ($\Phi^\dag$) and characterized by a K\"ahler potential $\Kcal(\Phi^\dag, \Phi)$, a superpotential, and a gauge kinetic function.
In this work, we consider a generic K\"ahler potential including higher-dimensional operators, but otherwise we assume the canonical gauge kinetic function since we focus only on gauge interactions, and we also ignore the corrections from superpotential since flavor observables in the first and second generations are sensitive to very high energy scale.

To obtain the one- and two-loop effective K\"ahler potential,
we expand around non-trivial background superfields $\phi$ and $\phi^\dag$: \textit{i.e.}, superfields are replaced as $\Phi \to \phi + \Phi$.

In the following notation, it is convenient to use the geometric structure for the generic K\"ahler potential.
The K\"ahler metric is defined as the second-derivative of the K\"ahler potential with respect to the background superfields,
\eqs{
G^a_{~b} \equiv \Kcal^a_{~b}
= \frac{\partial^2}{\partial \phi^\dag_a \partial \phi^b} \Kcal \, .
}
Here, subscripts and superscripts in the geometric quantity denote derivatives with respect to chiral and anti-chiral background superfields, respectively, and
$(G^{-1})^a_b$ indicates the inverse of the K\"ahler metric.
Given the metric, we construct the geometric quantities like a connection and a curvature.
In particular, the curvature is given by
\eqs{
R^{a~b~}_{~c~d} \equiv \Kcal^{ab}_{cd} - \Kcal^{ab}_{~~e} (G^{-1})^e_f \Kcal^{~~f}_{cd} \, .
}

In the background field method, gauge symmetries are also broken by the background superfields, and the masses for superfields depend on them.
The mass matrix for vector superfields is given by
\eqs{
(M_{V(\alpha,\beta)}^2 )_{IJ} &
= \frac12 \left((M_{C(\alpha,\beta)}^2 )_{IJ} + (M_{C(\beta,\alpha)}^2 )_{JI} \right) \, ,
}
with the mass matrix for ghost superfields,
\eqs{
(M_{C(\alpha,\beta)}^2 )_{IJ} & = 2 g_\alpha g_\beta \phi^\dag_a (\TIa)_b^a G^b_c (\TJb)_d^c \phi^d \, .
}
Here, $\alpha$ and $\beta$ indicate labels of gauge groups, and $I$ and $J$ represent the adjoint indices. Generators of the gauge group $\alpha$ are given by $\TIa$.
Summation about indices  of superfields $\phi$ and $\phi^\dag$ is implicit.
The mass matrix for chiral and anti-chiral superfields is
\eqs{
(M^2)^b_a
= 2 g_\alpha^2 (\TIa \phi)^b (\phi^\dag \TIa G)_a \, , \\
}
Here, summation about $\alpha$ is implicit again.

The effective K\"ahler potential is given using the above mass matrices, and thus it is a function of background superfields.
The one-loop effective K\"ahler potential is
\eqs{
\Kcal_{1L} = - \frac{1}{16\pi^2} \Tr ~ M_C^2 \left( 2 - \ln \frac{M_C^2}{\ovl \mu^2} \right) \, .
}
Here, the symbol ``$\Tr$'' denotes the trace of gauge adjoint indices $I, J, \cdots$.
We define $\ovl \mu \equiv \mu e^{- 4\pi \gamma}$, where $\mu$ denotes the $\ovl{\text{MS}}$ renormalization scale and $\gamma$ is the Euler number.

The two-loop effective K\"ahler potential is given by
\eqs{
& \Kcal_{2L}
= \frac12 R^{a~b~}_{~c~d} \ovl J^{c~d~}_{~a~b}(M^2)
- (G \TIa \phi)^c_{~;b} (\phi^\dag T^{(\alpha)}_J G)^{~;d}_a \ovl I^{a~b~IJ}_{~c~d}(M^2, M^2,M_V^2) \\
& - f^{(\alpha)}_{LIN} f^{(\alpha)}_{JKM} \ovl I^{IJKLMN} (M_V^2, M_V^2, M_V^2) \, .
\label{eq:univ_2loop}
}
Here, functions $\ovl I$ and $\ovl J$ denote loop integrals, which are shown in \cref{eq:loop_func}.
Structure constants $f^{(\alpha)}_{IJK}$ are defined by $[T^{(\alpha)}_I, T^{(\alpha)}_J] = i f^{(\alpha)}_{IJK} T^{(\alpha)}_K$.
The second term depends on covariant derivatives of $(G\TIa\phi)^a$ and $(\phi^\dag \TIa G)_a$:
\eqs{
(G\TIa\phi)^a_{~;b} &
= G^a_c (\TIa)^c_b + \Kcal^a_{bc} (\TIa\phi)^c \, , \\
(\phi^\dag \TIa G)_a^{~;b} &
=  (\TIa)^a_c G^c_b + (\phi^\dag \TIa)_c \Kcal_a^{bc} \, .
}
In \cref{eq:univ_2loop}, several terms in the original paper \cite{Nibbelink:2005wc} vanish since
we do not include the corrections from superpotential and we also assume the canonical gauge kinetic function.

\section{Two-loop Anomalous Dimensions\label{sec:anomalous}}
In this section, we derive two-loop anomalous dimensions for higher-dimensional K\"ahler potential terms.
As mentioned above, we only consider two-loop contributions via gauge interactions.
We employ the dimensional reduction ($\ovl{\text{DR}}$) scheme for the regularization.
To begin with, we show a derivative of the effective K\"ahler potential with respect to the renormalization scale $\mu$ in order to derive the anomalous dimension for the higher-dimensional operators.

We consider a generic supersymmetric model with a product gauge group $G = G_1 \times G_2 \times \cdots $.
We divide the tree-level K\"ahler potential as follows:
\eqs{
\Kcal & = \Kcal_0 + \Del \Kcal \, , \\
\Kcal_0 & = \sum_\Phi \Phi^\dag_a \left( \prod_\alpha e^{2g_\alpha V^I_\alpha \TIa} \right)^a_b \Phi^b \, , ~~~~~
\Del \Kcal = \mathcal{C} \Ocal + \mathcal{C} \Ocal^\dag \, ,
\label{eq:treeKahlar}
}
with the Wilson coefficient $\mathcal{C}$ of the operator,
\eqs{
\Ocal = (\ovl\lam_A^{a_1 \cdots a_m} \Phi^\dag_{a_1} \cdots \Phi^\dag_{a_m} ) \left(\prod_\alpha e^{2g_\alpha V^I_\alpha T^{(\alpha)}_{I:\mathrm{comp.}}}\right)^A_B (\lam^B_{b_1 \cdots b_n} \Phi^{b_1} \cdots \Phi^{b_n} ) \, .
}
The coefficient $\lam^B_{b_1 \cdots b_n}$ ($\ovl\lam_A^{a_1 \cdots a_m}$) is for the invariant tensor which makes a set of chiral (anti-chiral) superfields an irreducible representation under $G$.
Here, $\{ a_1, \cdots, a_m\}$ and $\{ b_1, \cdots, b_n\}$ denote gauge indices for constituent chiral superfields, and $A, B$ denote gauge indices for the composite operators.
We introduce generators of the gauge group $G_\alpha$ for the composite operator $\lam^B_{b_1 \cdots b_n} \Phi^{b_1} \cdots \Phi^{b_n}$ as $T^{(\alpha)}_{I:\mathrm{comp.}}$.
Note that $\Phi, \Phi^\dag$, and $V$ denote the quantum superfields while $\phi$ and $\phi^\dag$ represent the background superfields, in the following analysis.

We discard the higher order terms of $\Del \Kcal$ in the effective K\"ahler potential since they are irrelevant to derivation of the anomalous dimensions for $\Del \Kcal$.
The geometric quantities are simplified in this approximation.
The K\"ahler metric and its inverse are easily found to be
\eqs{
G^a_b = \del^a_b + (\Del \Kcal)^a_b \, , ~~~~~~
(G^{-1})^a_b = \del^a_b - (\Del \Kcal)^a_b \, ,
}
and the K\"ahler curvature is just the fourth derivative of $\Del\Kcal$,
\eqs{
R^{a~b~}_{~c~d} = (\Del \Kcal)^{ab}_{cd} \, .
}

We divide all mass matrices into canonical and $\Del \Kcal$ contributions.
First, for the ghost mass matrix, we have
\eqs{
(M_{C(\alpha,\beta)}^2)_{IJ}
& \equiv (M_{C0(\alpha,\beta)}^2 )_{IJ} + (\Del M_{C(\alpha,\beta)}^2 )_{IJ} \, ,
}
with
\eqs{
(M_{C0(\alpha,\beta)}^2)_{IJ}
& \equiv 2 g_\alpha g_\beta (\phi^\dag \TIa)_a (\TJb \phi)^a \, , \\
(\Del M_{C(\alpha,\beta)}^2 )_{IJ}
& \equiv 2 g_\alpha g_\beta (\phi^\dag \TIa)_a (\Del \Kcal)^a_b (\TJb \phi)^b \, .
}
Since we assume the vanishing superpotential and the renormalizable gauge kinetic function, the mass matrix for the chiral superfields is expanded as follows:
\eqs{
(M^2)^b_a & \equiv (M^2_0)^b_a + (\Del M^2)^b_a \, ,
}
with
\eqs{
(M^2_0)^b_a & \equiv 2 g_\alpha^2 (\TIa \phi)^b (\phi^\dag \TIa)_a \, , ~~~~~
(\Del M^2)^b_a \equiv 2 g_\alpha^2 (\TIa \phi)^b (\phi^\dag \TIa \Del\Kcal)_a \, .
}

Now, we expand the two-loop effective K\"ahler potential \cref{eq:univ_2loop} with respect to the higher-dimensional term $\Del \Kcal$.
The loop integrals $\ovl I$ and $\ovl J$ contain not only terms proportional to $\ln \mu$, but also constant and higher logarithmic terms.
We handle only the  $\ln \mu$ terms since only those terms are needed for derivation of the two-loop anomalous dimensions.

Before we show the $\ln \mu$ derivative of the two-loop K\"ahler potential, we subtract the $\ln \mu$ terms from the second term of \cref{eq:univ_2loop}.
This term is more complicated than other terms due to the non-Hermitian property of the mass matrix $M^2$.
We define the square roots of the K\"ahler metric as
\eqs{
G^a_b = \ovl E^a_x E^x_b \, , ~~~~~
(E^x_a)^\ast = \ovl E^a_x \, .
}
We write the propagators for chiral supermultiplets with the Hermitian mass matrix using the square roots as follows:
\eqs{
\Del_{\Phi^\dag \Phi} =
[\square - M^2]^{-1} G^{-1}
= E^{-1} [\square - M'^2]^{-1} \ovl E^{-1} \, ,
}
with the Hermitian mass matrix
\eqs{
(M'^2)_b^a
\equiv 2 g_\alpha^2 (E \TIa \phi)^a  (\phi^\dag \TIa \ovl E)_b \, .
}
We reconsider the second term of \cref{eq:univ_2loop} with the Hermitian mass matrix $M'^2$.
The loop function $\ovl I$ are changed due to the redefinition of the propagator and mass matrix:
\eqs{
\ovl I^{a~b~IJ}_{~c~d}(M^2, M^2, M_V^2)
\to (E^{-1})^a_x (E^{-1})^b_y \, \ovl I'^{x~y~IJ}_{~c~d}(M'^2, M'^2, M_V^2) \, .
}
The square roots $E^{-1}$ acting on the loop function arise from the redefinition of the propagator.
The prime of the redefined loop function represents that $G^{-1}$ in the loop function $\ovl I$ is replaced with $\ovl E^{-1}$.
The $\ln \mu$ part of the second term in \cref{eq:univ_2loop} is given by
\eqs{
& - (G \TIa \phi)^c_{~;b} (\phi^\dag \TJb G)^{~;d}_a (E^{-1})^a_x  (E^{-1})^b_y \left. \ovl I'^{x~y~IJ}_{~c~d}(M'^2, M'^2,M_V^2)\right|_{\ln \mu} \\
& = \frac{2 \ln  \mu^2}{(16\pi^2)^2} g_\alpha^2 g_\beta^2 \left\{
\del_{\alpha\beta} S_\alpha C_\alpha^{\mathrm{comp.}}\Del\Kcal \right. \\
& + \left[
(\phi^\dag \TJb \TIa \TIa)_a (\Del \Kcal)^a_b (\TJb \phi)^b
+ (\phi^\dag \TJb)_a (\Del \Kcal)^a_b (\TIa \TIa \TJb \phi)^b \right]\\
& + \phi^\dag(\TIa \TJb + \TJb \TIa) \phi ~ \tr \left[ \TJb (\Del \Kcal \TIa \phi) + \TIa (\Del \Kcal \phi^\dag \TJb) \right] \\
& + 2 (\phi^\dag \TJb)_a \left[ \Del\Kcal \TIa \phi + \Del\Kcal \phi^\dag \TIa \right]^a_b (\TIa \TJb \phi)^b \\
& \left. + 2 (\phi^\dag \TJb \TIa)_a \left[ \Del\Kcal \TIa \phi + \Del\Kcal \phi^\dag \TIa \right]^a_b (\TJb \phi)^b \right\}\, .
\label{eq:2ndterm}
}
Here, we define the sum of Dynkin indices for chiral supermultiplets as $S_\alpha \equiv \sum_{\phi} I_\alpha(\phi)$.
The quadratic Casimir invariant
$C_\alpha^{\mathrm{comp.}}$ is  for the composite chiral superfield
$\lam^B_{b_1 \cdots b_n} \Phi^{b_1} \cdots \Phi^{b_n}$ in $\Ocal$.
The following identity is derived using the relation between gauge transformations for constituent superfields $\phi$ and composite operators
\cite{Hisano:2013ege},
\eqs{
(\phi^\dag \TIa)_a (\Del \Kcal)^a_b (\TIa \phi)_b
= C_\alpha^{\mathrm{comp.}}\Del\Kcal \, .
\label{eq:quadcomp}
}
In the last three lines of \cref{eq:2ndterm}, we use the following short-hand notation,
\eqs{
(\Del \Kcal \TIa \phi)^a_b = (\Del\Kcal)^a_{bc} (\TIa \phi)^c \, , ~~~~~~
(\Del \Kcal \phi^\dag \TIa)^a_b =  (\Del\Kcal)^{ac}_b (\phi^\dag \TIa)_c \, .
}

As a result, we obtain the $\ln \mu$ derivative of the two-loop effective K\"ahler potential as follows:
\eqs{
&(16\pi^2)^2 \mu \pdiff{1}{}{\mu} \Kcal_{2L} \\
& = \sum_\alpha 4 (S_\alpha -3 C_\alpha(\mathrm{Ad.})) C_\alpha^{\mathrm{comp.}} g_\alpha^4 ~ \Del \Kcal \\
& + \sum_{\alpha, \beta} 4 g_\alpha^2 g_\beta^2  \left\{ \left[
(\phi^\dag \TJb \TIa \TIa)_a (\Del \Kcal)^a_b (\TJb \phi)^b
+ (\phi^\dag T_J)_a (\Del \Kcal)^a_b (\TIa \TIa \TJb \phi)^b
\right] \right.\\
& + \left[ \sum_\phi \phi^\dag(\TIa \TJb + \TJb \TIa) \phi \right]~ \tr \left[ \TJb (\Del \Kcal \TIa \phi) + \TIa (\Del \Kcal \phi^\dag \TJb) \right] \\
& + 2 (\phi^\dag \TJb)_a \left[
\Del\Kcal \TIa \phi + \Del\Kcal \phi^\dag \TIa
\right]^a_b (\TIa \TJb \phi)^b \\
& + 2 (\phi^\dag \TJb \TIa)_a \left[
\Del\Kcal \TIa \phi + \Del\Kcal \phi^\dag \TIa
\right]^a_b (\TJb \phi)^b \\
& \left. + 2 (\Del \Kcal)^{ab}_{cd} (\phi^\dag \TIa)_a (\phi^\dag \TJb)_b (\TIa \phi)^c (\TJb \phi)^d \right\} \, .
\label{eq:2Loop_Kahler}
}
Here, we show only the first-order terms with respect to $\Del \Kcal$,
and the logarithmic terms are neglected.
The last term comes from the curvature term in \cref{eq:univ_2loop}, and the term proportional to $C_\alpha(\mathrm{Ad.})$, which is the quadratic Casimir invariant for the adjoint representation,  arises from the third term in \cref{eq:univ_2loop}.

Next, we derive the anomalous dimensions for higher-dimensional operators $\Ocal$ using the RGEs for vertex functions.
The RGEs for the vertex functions with an insertion of the operator $\Ocal$ are given by
\eqs{
\left[ \mu \pdiff{1}{}{\mu}
+ \sum_\alpha \beta_\alpha \pdiff{1}{}{g_\alpha}
- \left(
\gamma_\phi \, \phi \pdiff{1}{}{\phi}
+ \gamma_{\phi^\dag} \, \phi^\dag \pdiff{1}{}{\phi^\dag}
\right)
+ \gamma_{\Ocal} \right] \Gamma_{\Ocal}
= 0 \, .
\label{eq:CSeq}
}
Here, $\beta_\alpha$ is the beta function for $g_\alpha$.
The sum of constituent superfields in the third term is implicit, and
$\gamma_\phi$ and $\gamma_{\phi^\dag}$ are, respectively, anomalous dimensions for the chiral superfield $\phi$ and anti-chiral superfield $\phi^\dag$ composing the operator $\Ocal$.
The one-loop anomalous dimensions $\gamma_\phi^{(1)}$ and two-loop anomalous dimensions $\gamma_\phi^{(2)}$ are, respectively,
\eqs{
\gamma_\phi^{(1)} & =
- 2 \sum_\alpha \frac{g_\alpha^2}{16\pi^2} C_\alpha(\phi) \, , \\
\gamma_\phi^{(2)} & =
2 \sum_{\alpha,\beta} \frac{g_\alpha^2 g_\beta^2}{(16\pi^2)^2} C_\alpha(\phi) \left[ b_\alpha \del_{\alpha,\beta} + 2 C_\beta(\phi) \right] \, .
}
We have already given the partial derivative of the vertex function with respect to the renormalization scale $\mu$ in \cref{eq:2Loop_Kahler}.

Now, we derive $\gamma_\Ocal$ that is the anomalous dimension for the operator $\Ocal$.
Using $\gamma_\phi^{(1)}$ and \cref{eq:CSeq} at one-loop order in the perturbation theory, we find the one-loop anomalous dimension of a generic higher-dimensional operator $\Ocal$ \cite{Hisano:2013ege},
\eqs{
\gamma_{\Ocal}^{(1)}
& = \sum_\alpha \frac{g_\alpha^2}{16\pi^2} \left[
4 C_\alpha^{\mathrm{comp.}}
- 2 \sum_\phi C_\alpha(\phi)
\right] \, .
}
Here, the summation symbol about $\phi$ represents the sum of all chiral and anti-chiral superfields composing the operator $\Ocal$, unless otherwise stated explicitly.

The anomalous dimension for $\Ocal$ includes the fourth derivative term which arises from the curvature term in \cref{eq:2Loop_Kahler}.
We can eliminate this term using the relation given by \cref{eq:quadcomp}.
When the second derivative, $(\phi^\dag T_J^{(\beta)})_a (T_J^{(\beta)} \phi)^b \partial^2/\partial \phi^\dag_a \partial \phi^b $, acts on the both sides of \cref{eq:quadcomp}, we obtain
\eqs{
& (\Del \Kcal)^{ab}_{cd} (\phi^\dag \TIa)_a (\phi^\dag \TJb)_b (\TIa \phi)^c (\TJb \phi)^d \\
& =C_\alpha^{\mathrm{comp.}} C_\beta^{\mathrm{comp.}} \Del\Kcal
- (\phi^\dag \TJb)_a (\Del\Kcal \phi^\dag \TIa)^a_b (\TIa \TJb \phi)^b \\
& - (\phi^\dag \TJb \TIa)_a (\Del\Kcal \TIa \phi)^a_b (\TJb \phi)^b
- (\phi^\dag \TJb \TIa)_a (\Del\Kcal)^a_b (\TIa \TJb \phi)^b \, .
}

Furthermore, if we constraint the form of the higher-dimensional operators, the two-loop anomalous dimensions are more simplified.
In this work, we concentrate on the K\"ahler potentials which include dimension-six four-Fermi operators,
\eqs{
\Del \Kcal = (\lam_A^{a_1 a_2} \phi^\dag_{a_1} \phi^\dag_{a_2}) (\lam^A_{b_1 b_2} \phi^{b_1}\phi^{b_2}) + \mathrm{h.c.} \, .
\label{eq:fourfermikahler}
}
This form can be applied to the operators for proton decay processes, flavor-changing processes, and so on.
We can reduce the two-loop anomalous dimensions in this setup.

The generic K\"ahler potential is invariant under the global symmetry $G$ when taking $V=0$.
It implies
\eqs{
\Kcal(\phi^\dag ~ e^{-i \alpha}, e^{i \alpha} \phi) - \Kcal(\phi^\dag , \phi) = 0 \, ,
\label{eq:symmetry}
}
where $\alpha = \alpha_I \TIa$ denotes the constant transformation parameter.
Each order of $\alpha$ should be vanished in \cref{eq:symmetry} when we expand the above identity in $\alpha$.
We obtain the following relation on the K\"ahler potential from the $\alpha^2$ term,
\eqs{
(\phi^\dag \TIa \TJb)_a \Kcal^a
+ (\TIa \TJb \phi)^a \Kcal_a
+ (\phi^\dag \TIa)_a (\phi^\dag \TJb)_b \Kcal^{ab}
+ (\TIa \phi)^a (\TJb \phi)^b \Kcal_{ab} \\
= 2 (\phi^\dag \TIa)_a \Kcal^a_b (\TJb \phi)^b \, .
\label{eq:gauge_inv}
}
We take the gauge groups as $\alpha = \beta$ and the gauge indices as $I = J$, and we sum up the gauge indices in both sides.
Besides, we operate $(\TJb \phi)^a (\phi^\dag \TJb)_b \partial^2/\partial \phi^a \partial \phi^\dag_b$ on both sides, and then we obtain
\eqs{
& \sum_\phi C_\alpha(\phi) C_\beta^{\mathrm{comp.}} \Del \Kcal \\
& + 2 (\phi^\dag \TJb \TIa)_a (\Del \Kcal \phi^\dag \TIa)_b^a (\TJb \phi)^b
+ 2 (\phi^\dag \TJb)_a (\Del \Kcal \TIa \phi )^a_b (\TIa \TJb \phi)^b \\
& = 2 C_\alpha^{\mathrm{comp.}} C_\beta^{\mathrm{comp.}} \Del \Kcal \, .
}
Here, we neglect the third derivative of $\Del\Kcal$ with respect to chiral superfields since we focus on the four-Fermi type K\"ahler potential \cref{eq:fourfermikahler}.
Only $\Del \Kcal$ terms appear in this relation since the canonical term satisfies \cref{eq:gauge_inv} trivially.

Using the above relations, we obtain the two-loop anomalous dimension for a generic dimension-six operators $\Ocal$ as follows:
\eqs{
& (16 \pi^2)^2\gamma^{(2)}_{\Ocal} \Ocal \\
& = 2\left[ \sum_\phi \left( b_\alpha C_\alpha(\phi) \del_{\alpha,\beta}
+ 2 C_\alpha(\phi) C_\beta(\phi) \right)
- \left( 4 C_\alpha^{\mathrm{comp.}} - 2 \sum_\phi C_\alpha(\phi) \right) C_\beta^{\mathrm{comp.}}
 \right] g_\alpha^2 g_\beta^2 \, \Ocal \\
& - 4 g_\alpha^2 g_\beta^2 \left[
(\phi^\dag \TJb \TIa \TIa)_a \Ocal^a_b (\TJb \phi)^b
+ (\phi^\dag \TJb)_a \Ocal^a_b (\TIa \TIa \TJb \phi)^b
\right] \\
& - 4 g_\alpha^2 g_\beta^2 \left[\sum_\phi \phi^\dag(\TIa \TJb + \TJb \TIa) \phi \right] ~ \tr \left[ \TJb (\Ocal \TIa \phi) + \TIa (\Del\Kcal \phi^\dag \TJb) \right] \\
& + 8 g_\alpha^2 g_\beta^2 (\phi^\dag \TIa \TJb)_a \Ocal^a_b (\TJb  \TIa \phi)^b \, .
\label{eq:2loopano}
}
Here, the sum about the gauge group $\alpha$ and $\beta$ is implicit, and $b_\alpha = S_\alpha - 3 C_\alpha(\mathrm{Ad.})$ is the one-loop coefficient of the beta function $\beta_\alpha$.
While the first term is completely determined by the group theoretical factors, the other terms cannot be simplified anymore.
Indeed, as we will see in the next section, the last three terms give rise to the operator mixing depending on the operators.

\section{Applications \label{sec:application}}

In this section, we discuss concrete examples and
derive two-loop anomalous dimensions for concrete operators.
In particular, we focus on $\Del F = 2$ flavor-violating operators and proton decay operators.
For these operators, the second line from the last in \cref{eq:2loopano} vanishes since it contains the summation of flavor indices.
If we consider the $\Del F=1$ flavor-violating processes such as rare meson decays, lepton decays, and $\mu$-$e$ conversion, this term should be taken into consideration.
For the proton decay operators, the next-leading order corrections, the two-loop RGE analysis \cite{Hisano:2013ege} and finite corrections at the grand unification scale \cite{Hisano:2015ala,Bajc:2016qcc}, have already been estimated.
However, the explicit derivations for the proton decay operators have not shown in Ref.~\cite{Hisano:2013ege}.
For the sake of completeness, we show the two-loop anomalous dimensions for the proton decay operators in this section.

In the following we classify the $\Del F = 2$ flavor-violating operators in the K\"ahler potential into two types.
The first ones are composed of chiral superfields which transform as a common representation under the SM gauge group, as $\Phi^\dag_i  \Phi^\dag_i  \Phi_j \Phi_j$ ($i,j$ are flavor indices and $i\ne j$).
Here, $\Phi$ represents the matter chiral superfields in the MSSM: quark doublet ($Q$), lepton doublet ($L$), up-type quark singlet ($\ovl U$), down-type quark singlet ($ \ovl D$), and charged lepton singlet ($\ovl E$).
We refer to them as $\Phi\Phi$ operators.
The second ones are composed of two different chiral superfields, such as $Q_2^\dag \ovl D_1^\dag Q_1 \ovl D_2$ and $Q_2^\dag \ovl U_1^\dag Q_1 \ovl U_2$.
We call them as $\Phi \Psi$ operators.

\subsection{$\Phi\Phi$ Operators}

First we consider the $\Phi\Phi$ operators.
As we have seen in \cref{eq:treeKahlar}, the chiral part of the K\"ahler potential should be decomposed into the irreducible components.
The K\"ahler potential for these operators is given by
\eqs{
\Del \Kcal = \sum_{\Pcal} C_{\Pcal} \Ocal^\Pcal \, , ~~~~
\Ocal^\Pcal \equiv \Phi^\dag_i  \Phi^\dag_i \Pcal \Phi_j \Phi_j \, .
}
Here, $\Pcal$s project out the irreducible components in the chiral part, and the gauge indices are implicit.
$C_{\Pcal}$ is the Wilson coefficient for the effective operator $\Ocal^\Pcal$.
Hereafter, we omit the flavor indices, $i$ and $j$, since we consider only the gauge interactions.

If the constituent chiral superfield $\Phi$ transforms as a fundamental representation under only a single non-Abelian gauge group $SU(N)$, $\Pcal$ projects a product of two $\Phi$s onto a symmetric or an antisymmetric part.
The symmetric and antisymmetric projection operators on an $SU(N)$ group are denoted by $S_N$ and $A_N$, respectively.
They are defined by
\eqs{
(S_N)^{ab}_{cd} & \equiv \frac12 (\del^a_c \del^b_d + \del^b_c \del^a_d) \, , \\
(A_N)^{ab}_{cd} & \equiv \frac12 (\del^a_c \del^b_d - \del^b_c \del^a_d) \, . \label{eq:projectionop}
}
Here, $a, b, \cdots$ are the $SU(N)$ gauge indices.
The subscripts in $S_N$ and $A_N$ denote the gauge group $SU(N)$.
When we consider the chiral superfield which transforms as a bi-fundamental representation of $SU(N) \times SU(M)$ such as $Q$ in the MSSM, there are four projection operators,
\eqs{
\Pcal = S_N S_M \, , ~~
S_N A_M \, , ~~
A_N S_M \, , ~~
A_N A_M \, .
}

As a result, we obtain the two-loop anomalous dimension for $\Ocal^\Pcal$ which consists of $SU(N)$ fundamental chiral superfields $\Phi$ with $U(1)$ charges $q_\phi$ as follows:
\eqs{
\gamma^{(2)}_{\Ocal^\Pcal} = \frac{g_N^4}{(16\pi^2)^2} [ \gamma^{(2)}_{\Ocal^\Pcal} ]_{NN}
+ \frac{g_N^2 g_1^2}{(16\pi^2)^2} [ \gamma^{(2)}_{\Ocal^\Pcal} ]_{N1}
+ \frac{g_1^4}{(16\pi^2)^2} [ \gamma^{(2)}_{\Ocal^\Pcal} ]_{11} \, ,
\label{eq:anomalousPhPh1}
}
with
\eqs{
[ \gamma^{(2)}_{\Ocal^\Pcal} ]_{NN} & =
8 \left[ b_N C_N + 2 (C_N)^2 - (C_N^{\mathrm{comp.}}-2C_N )C_N^{\mathrm{comp.}} \right] \\
& - 4 \left[ \left(c_\Pcal - \frac{4}{N} \right) C_N - \frac{1}{2N} \left( c_\Pcal (N^2 - 2) + \frac{4}{N} \right)\right] \, , \\
[ \gamma^{(2)}_{\Ocal^\Pcal} ]_{11} & =
8 \left[ b_1 q_\phi^2 + 2 q_\phi^4 - (\qcomp^2-2 q_\phi^2) \qcomp^2 \right] \, , \\
[ \gamma^{(2)}_{\Ocal^\Pcal} ]_{N1} & =
- 4 \left[ 4 \CNcomp \qcomp^2 -4 C_N \qcomp^2 - 4 q_\phi^2 \CNcomp - q_\phi^2 \left(c_\Pcal - \frac{4}{N} \right) \right]
\, . \\
}
Here, $b_1$ and $b_N$ are respectively the one-loop coefficients of gauge couplings $g_1$ for $U(1)$ and $g_N$ for $SU(N)$.
$q_{\mathrm{comp.}}$ is the $U(1)$ charge of the composite operator, and thus $q_{\mathrm{comp.}} = 2 q_\phi$.
The constant $c_\Pcal$ depends on the projection operator $\Pcal$, and it is given as
\eqs{
c_\Pcal =
\begin{cases}
  2(N+1) & (\Pcal = S_N) \, , \\
  2(N-1) & (\Pcal = A_N) \, . \\
\end{cases} \label{eq:cPN}
}

In the case of the bi-fundamental chiral superfield of $SU(N) \times SU(M)$, the two-loop anomalous dimension for the effective operator $\Ocal^\Pcal$ is decomposed as follows:
\eqs{
\gamma^{(2)}_{\Ocal^\Pcal} \Ocal^\Pcal = & ~
\frac{g_N^4}{(16\pi^2)^2} [ \gamma^{(2)}_{\Ocal^\Pcal} \Ocal^\Pcal ]_{NN}
+ \frac{g_M^4}{(16\pi^2)^2} [ \gamma^{(2)}_{\Ocal^\Pcal} \Ocal^\Pcal ]_{MM}
+ \frac{g_1^4}{(16\pi^2)^2} [ \gamma^{(2)}_{\Ocal^\Pcal} \Ocal^\Pcal ]_{11} \\
& + \frac{g_N^2 g_M^2}{(16\pi^2)^2} [ \gamma^{(2)}_{\Ocal^\Pcal} \Ocal^\Pcal ]_{NM}
+ \frac{g_N^2 g_1^2}{(16\pi^2)^2} [ \gamma^{(2)}_{\Ocal^\Pcal} \Ocal^\Pcal ]_{N1}
+ \frac{g_M^2 g_1^2}{(16\pi^2)^2} [ \gamma^{(2)}_{\Ocal^\Pcal} \Ocal^\Pcal ]_{M1} \, ,
\label{eq:anomalousPhPh2}
}
with the $SU(M)$ coupling $g_M$.
In this case, there is mixing among the operators in the $NM$ mixed term.
The constant $c_\Pcal$ in the $NN$ and $N1$ components are replaced with $c_{P_N}$ since the projection operator $\Pcal$ is a product of projections $P_N$ and $P_M$ for $SU(N)$ and $SU(M)$ indices: $\Pcal = P_N P_M$.
The $MM$ and $M1$ components of the matrix are respectively obtained by replacing $N$ with $M$ in the $NN$ and $N1$ components.
The $SU(N)$-$SU(M)$ mixing term is obtained as follows:
\eqs{
[ \gamma^{(2)}_{\Ocal^\Pcal} \Ocal^\Pcal]_{NM}
& = [\gamma^{(2)}_{\Ocal^\Pcal}]^{\Pcal\Pcal}_{NM} \Ocal^\Pcal + [\gamma^{(2)}_{\Ocal^\Pcal}]^{\Pcal\ovl{\Pcal}}_{NM} {\Ocal}^{\ovl{\Pcal}} \, ,
}
with coefficients,
\eqs{
[\gamma^{(2)}_{\Ocal^\Pcal}]^{\Pcal\Pcal}_{NM} = & ~
16(C_N C_M
- \CNcomp \CMcomp
+ C_N \CMcomp
+ C_M \CNcomp ) \\
& + \left(c_{P_N} - \frac{4}{N} - 4 C_N \right)
\left(c_{P_M} - \frac{4}{M} - 4 C_M \right) \, , \\
[\gamma^{(2)}_{\Ocal^\Pcal}]^{\Pcal\ovl{\Pcal}}_{NM}
= & ~ c_{P_N} c_{P_M} \, .
}
Here, the projection operator with an overline, $\ovl\Pcal = \ovl P_N \ovl P_M$, is defined as $\ovl S_N = A_N$ and $\ovl A_N = S_N$.

In the MSSM, there is the composite operator of which constituent superfields have different $U(1)$ charges; $\ovl U^\dag \ovl D^\dag \ovl U \ovl D$.
For this type of operators, the K\"ahler potential is given by
\eqs{
\Del \Kcal = \sum_{\Pcal} C_{\Pcal} \Ocal^\Pcal \, , ~~~~
\Ocal^\Pcal \equiv \Phi^\dag_1  \Phi^\dag_2 \Pcal \Phi_1 \Phi_2 \, ,
}
with $U(1)$ charges $q_1$ and $q_2$ for $\Phi_1$ and $\Phi_2$, respectively.
We assume that the $\Phi_i$ $(i = 1, 2)$ transforms as a (anti-)fundamental representation of $SU(N)$.
While the $NN$-component of the two-loop anomalous dimension is the same as \cref{eq:anomalousPhPh1}, the $11$-component and the $N1$-component are modified as follows:
\eqs{
[ \gamma^{(2)}_{\Ocal^\Pcal}]_{11}
& =
4 b_1 (q_1^2 + q_2^2)
+ 8 (q_1^4 + q_2^4)
- 8 [\qcomp^2 - (q_1^2+q_2^2)]\qcomp^2
- 8 q_1 q_2 (q_1-q_2)^2 \, , \\
}
with $\qcomp = q_1 + q_2$, and
\eqs{
[ \gamma^{(2)}_{\Ocal^\Pcal} \Ocal^\Pcal ]_{N1}
& =
[\gamma^{(2)}_{\Ocal^\Pcal}]^{\Pcal\Pcal}_{N1} \Ocal^\Pcal + [\gamma^{(2)}_{\Ocal^\Pcal}]^{\Pcal\ovl{\Pcal}}_{N1} {\Ocal}^{\ovl{\Pcal}} \, ,
}
with
\eqs{
[\gamma^{(2)}_{\Ocal^\Pcal}]^{\Pcal\Pcal}_{N1}
= & ~ 8 C_N \left[ 2 (q_1^2 + q_2^2) - (q_1 + q_2)^2 \right] + 4 q_1 q_2 \left(c_\Pcal - \frac{4}{N} \right) \\
& - 16 \CNcomp \qcomp^2
+ 16 C_N \qcomp^2
+ 8 (q_1^2 + q_2^2) \CNcomp \, , \\
[\gamma^{(2)}_{\Ocal^\Pcal}]^{\Pcal\ovl{\Pcal}}_{N1}
= & ~ 2 c_\Pcal (q_1 - q_2)^2
\, . \\
}

\subsection{$\Phi\Psi$ Operators}

Now, we consider the composite operators in $\Del \Kcal$ which consist of chiral superfields $\Phi$ and $\Psi$ transforming differently under gauge symmetry, $SU(N) \times SU(M) \times U(1)$.
The charge assignments of $\Phi$ and $\Psi$ of the gauge symmetry are assumed to be
\eqs{
\Phi: (\mathbf{N},\mathbf{M})_{q_\phi} \, , ~~~~~
\Psi: (\ovl{\mathbf{N}},\mathbf{1})_{q_\psi} \, ,
}
Here, $q_{\phi}$ and $q_{\psi}$ denote $U(1)$ charges for $\Phi$ and $\Psi$, respectively.
$\Phi$ ($\Psi$) is a fundamental (anti-fundamental) representation of $SU(N)$ while only $\Phi$ is charged under $SU(M)$.

As mentioned in the previous subsection, it is necessary to use the decomposition of the chiral part in the effective operator into the irreducible components.
The K\"ahler potential consisting of these chiral superfields is given by
\eqs{
\Del \Kcal = \sum_{i=1}^2 C^{(i)} \Ocal^{(i)} \, ,
}
where $C^{(i)}$ is the Wilson coefficient and the irreducible operators are
\eqs{
\Ocal^{(1)} \equiv \Ocal_{\mathrm{adj.}}
& = 2 (\Phi^\dag \TIa \Psi^\dag) (\Psi \TIa \Phi) \, , \\
\Ocal^{(2)} \equiv \Ocal_{\mathrm{sing.}}
& = (\Phi^\dag \Psi^\dag) (\Psi \Phi) \, .
\label{eq:Operator_phipsi}
}
Here, we omit the flavor indices, again.
Brackets indicate the contraction with respect to the $SU(N)$ indices, and $\TIa$ stands for the $SU(N)$ generator.
The chiral part of $\Ocal^{(1)}$ transforms as the adjoint representation under the $SU(N)$ while that of $\Ocal^{(2)}$ does as the singlet representation.

As is the case with the previous subsection, we decompose the two-loop anomalous dimensions for these operators as follows:
\eqs{
\gamma^{(2)}_{ij} \Ocal^{(j)} = & ~
\frac{g_N^4}{(16\pi^2)^2} [ \gamma^{(2)}_{ij} ]_{NN} \Ocal^{(j)}
+ \frac{g_M^4}{(16\pi^2)^2} [ \gamma^{(2)}_{ij} ]_{MM} \Ocal^{(j)}
+ \frac{g_1^4}{(16\pi^2)^2} [ \gamma^{(2)}_{ij} ]_{11} \Ocal^{(j)} \\
& + \frac{g_N^2 g_M^2}{(16\pi^2)^2} [ \gamma^{(2)}_{ij} ]_{NM} \Ocal^{(j)}
+ \frac{g_N^2 g_1^2}{(16\pi^2)^2} [ \gamma^{(2)}_{ij} ]_{N1} \Ocal^{(j)}
+ \frac{g_M^2 g_1^2}{(16\pi^2)^2} [ \gamma^{(2)}_{ij} ]_{M1} \Ocal^{(j)} \, .
\label{eq:anomalousPhPs2}
}
Using \cref{eq:2loopano}, we obtain the two-loop anomalous dimensions for the $\Phi \Psi$ effective operators.
In this case, the effective operators mix with each other via the $SU(N)$ interactions at the two-loop level.
First, the terms in $\gamma_{ij}^{(2)}$ without operator mixing are obtained as
\eqs{
[\gamma^{(2)}]_{MM} = & ~
4 \left[b_M C_M + 2 (C_M)^2 \right] \mathbf{1} \, , \\
[\gamma^{(2)}]_{11} = & ~
4 \left[ b_1 (q_\phi^2 + q_\psi^2) + 2 \left(q_\phi^2 -4 q_\phi q_\psi + q_\psi^2 \right) \left(q_\phi^2 + q_\phi q_\psi + q_\psi^2 \right) \right] \mathbf{1} \, , \\
[\gamma^{(2)}]_{M1} = & ~
8 C_M q_\phi \left(2 q_\phi - 3 q_\psi \right) \mathbf{1} \, .
}
Here, $\mathbf{1}$ denotes a $2 \times 2$ unit matrix.
In this derivation, we use the fact that the chiral part of each effective operators transforms as the fundamental representation under the $SU(M)$.
That is to say, the quadratic Casimir invariant for the effective operator satisfies $\CMcomp = C_M(\phi)$.

Second, we show the terms in $\gamma^{(2)}_{ij}$ with operator mixing.
We find $[\gamma^{(2)}]_{NN}$ as follows:
\eqs{
[\gamma^{(2)}_{11}]_{NN}
& = 8 \left[ b_N C_N +2(C_N)^2 - \CNcomp(\CNcomp - 2 C_N ) - \frac{C_N}{N} + \frac{5}{2N^2} \right] \, , \\
[\gamma^{(2)}_{12}]_{NN}
& = \frac{4(N^2-1)}{N^2} \left( 2 C_N - \frac{3}{N} \right) \, , \\
[\gamma^{(2)}_{21}]_{NN}
& = \frac{4(N^2-4)}{N} \, , \\
[\gamma^{(2)}_{22}]_{NN}
& = 8 \left[b_N C_N +2(C_N)^2 + \frac{N^2-1}{N^2} \right] \, . \\
}
Here, $\CNcomp$ represents the quadratic Casimir invariant for the chiral part of $\Ocal^{(1)}$, that is $\CNcomp = N$ in this case.
The chiral part of $\Ocal^{(2)}$ transforms as a singlet representation of $SU(N)$, and thus the quadratic Casimir invariant is equal to zero.
$N$ appears explicitly when we calculate the second and the fourth lines in \cref{eq:2loopano}.
It is difficult to associate these terms with the group theoretical factors in our procedure.

The anomalous dimensions that arise from $SU(N)$-$SU(M)$ and $SU(N)$-$U(1)$ interactions are given by
\eqs{
[\gamma^{(2)}_{11}]_{NM}
& = 8 C_M \left( 4 C_N - \CNcomp - \frac{3}{2N} \right) \, , \\
[\gamma^{(2)}_{12}]_{NM}
& = 4 C_M \left( \frac{N^2-1}{N^2} + \frac{2}{N} C_N \right) \, , \\
[\gamma^{(2)}_{21}]_{NM}
& = 8 C_M \, , \\
[\gamma^{(2)}_{22}]_{NM}
& = 24 C_M C_N \, , \\
[\gamma^{(2)}_{11}]_{N1}
& = 16 \left[ \left(3 C_N - \CNcomp - \frac{3}{4N} \right) (q_\phi^2 + q_\psi^2)
+ \left(3 C_N - 4 \CNcomp - \frac{2}{N} \right) q_\phi q_\psi
\right] \, , \\
[\gamma^{(2)}_{12}]_{N1}
& = 16 \left[ \left(\frac{C_N}{2N} + \frac{N^2-1}{4N^2} \right) (q_\phi^2 + q_\psi^2)
- \frac{N^2-1}{N^2} q_\phi q_\psi
\right] \, , \\
[\gamma^{(2)}_{21}]_{N1}
& = 8 (q_\phi-q_\psi)^2 \, , \\
[\gamma^{(2)}_{22}]_{N1}
& = 8 C_N \left[ 5 (q_\phi^2 + q_\psi^2) + 6 q_\phi q_\psi \right] \, . \\
}
Since only $\Phi$ is charged under $SU(M)$, the group theoretical factor, $C_M$, is factored out.

Note that, in the MSSM, this type of operators includes not only the following effective K\"ahler operators,
\eqs{
Q^\dag \ovl U^\dag Q \ovl U \, , ~~~~~
Q^\dag \ovl D^\dag Q \ovl D \, , ~~~~~
Q^\dag L^\dag Q L \, ,
}
but also the following K\"ahler operators,
\eqs{
&
Q^\dag \ovl E^\dag Q \ovl E \, , ~~~~~
\ovl U^\dag  L^\dag \ovl U L \, , ~~~~~
\ovl U^\dag \ovl E^\dag \ovl U \ovl E \, , ~~~~~ \\
&
\ovl D^\dag  L^\dag \ovl D L \, , ~~~~~
\ovl D^\dag \ovl E^\dag \ovl D \ovl E \, , ~~~~~
L^\dag \ovl E^\dag L \ovl E \, .
}
Indeed, it is sufficient to calculate $[\gamma^{(2)}]_{MM}, [\gamma^{(2)}]_{11}$, and $[\gamma^{(2)}]_{M1}$ for the latter operators.
This is because that one of constituent chiral superfields of them transforms as non-trivial representation under given non-Abelian gauge group, and therefore the chiral part of the latter operators transforms as a fundamental representation under the SM gauge group.

\subsection{Proton Decay Operators}
In the last of this section, we show the two-loop anomalous dimensions for proton decay operators.
The effective K\"ahler term is given by
\eqs{
\Del \Kcal = \sum_{i=1}^2 C^{(i)}\Ocal^{(i)} + \text{h.c.} \, ,
\label{eq:Dim6-OperatorsSUSY}
}
with operators
\eqs{
\Ocal^{(1)} & = \ep_{\alpha\beta\gamma} \ep_{rs} (\ovl U^\dag)^\alpha (\ovl D^\dag)^\beta Q^{r\gamma} L^s \, , \\
\Ocal^{(2)} & = \ep_{\alpha\beta\gamma} \ep_{rs} \ovl E^\dag (\ovl U^\dag)^\alpha  Q^{r\beta} Q^{s\gamma} \, , \\
}
and $C^{(i)}$ ($i=1,2$) corresponds to the Wilson coefficients.

We divide the two-loop anomalous dimension into the following parts:
\eqs{
[\gamma_{\Ocal^{(i)}}^{(2)}] = & ~
\frac{g_3^4}{(16\pi^2)^2} [\gamma_{\Ocal^{(i)}}^{(2)}]_{33}
+ \frac{g_2^4}{(16\pi^2)^2} [\gamma_{\Ocal^{(i)}}^{(2)}]_{22}
+ \frac{g_Y^4}{(16\pi^2)^2} [\gamma_{\Ocal^{(i)}}^{(2)}]_{YY} \\
& + \frac{g_3^2 g_2^2}{(16\pi^2)^2} [\gamma_{\Ocal^{(i)}}^{(2)}]_{32}
+ \frac{g_3^2 g_Y^2}{(16\pi^2)^2} [\gamma_{\Ocal^{(i)}}^{(2)}]_{3Y}
+ \frac{g_2^2 g_Y^2}{(16\pi^2)^2} [\gamma_{\Ocal^{(i)}}^{(2)}]_{2Y} \, .
}
Since the chiral part of both operators transforms as a color triplet and a weak singlet under the SM gauge group, we are able to handle them simultaneously.
Even for these operators, what we should do is to calculate the second and fourth terms in \cref{eq:2loopano}.
We omit the details of the calculation, which is rather straightforward.
We obtain $[\gamma_{\Ocal^{(i)}}^{(2)}]_{\alpha\beta}$ symbolically as follows:
\eqs{
[\gamma_{\Ocal^{(i)}}^{(2)}]_{33} & = 6 C_3 (b_3+2C_3)
+ \frac{64}{9} - 4 C_3^{\text{comp.}}(2C_3^{\text{comp.}}-C_3) \, , \\
[\gamma_{\Ocal^{(i)}}^{(2)}]_{22} & = 4 C_2 (b_2+2C_2) \, , \\
[\gamma_{\Ocal^{(i)}}^{(2)}]_{YY}
& = 2b_Y S_{q^2}
+ 4 S_{q^4}
- 2 q_{\text{comp.}}^2(4 q_{\text{comp.}}^2
- 2 S_{q^2}) \\
& - 4 (\mathcal{S}_{1,3} + \mathcal{S}_{3,1} -2 \mathcal{S}_{2,2}) \, , \\
[\gamma_{\Ocal^{(i)}}^{(2)}]_{32}
& = 8 \sum_\phi C_3(\phi) C_2(\phi) + 4 C_2 C_3^{\text{comp.}} \, ,\\
[\gamma_{\Ocal^{(i)}}^{(2)}]_{2Y}
& = 8 S_{W:q^2}
+ 8 C_2 q_{\text{comp.}}^2
- 4 q_{\mathrm{comp.}} S_{W:q} \, ,\\
[\gamma_{\Ocal^{(i)}}^{(2)}]_{3Y}
& = 8 S_{C:q^2}
- 16 q_{\text{comp.}}^2 C_3^{\text{comp.}}
+ 12 q_{\text{comp.}}^2 C_3
+ 4 C^{\text{comp.}}_3 S_{q^2} \\
& -4 q_{\mathrm{comp.}} S_{C:q}
- 4 (\mathcal{S}_{C:0,2} + \mathcal{S}_{C:2,0} - 4 \mathcal{S}_{C:1,1}) \, .
\label{eq:symb_pdecay}
}
Here, $q_{\mathrm{comp.}}$ denotes the $U(1)$ hypercharge of the chiral part of the effective operator, $q_{\mathrm{comp.}} = -1/3$ for $\Ocal^{(1)}$ and $q_{\mathrm{comp.}} = 1/3$ for $\Ocal^{(2)}$.
The definitions of $S_{q^n}, S_{W:q^n}$, and $S_{C:q^n}$ are given by
\eqs{
S_{q^n} \equiv \sum_{\phi} q_\phi^n \, , ~~~~~
S_{W:q^n} \equiv \sum_{\phi} C_2(\phi) q_\phi^n \, , ~~~~~
S_{C:q^n} \equiv \sum_{\phi} C_3(\phi) q_\phi^n \, .
}
Here, we take the sum over all the chiral and anti-chiral superfields.
The definitions of $\mathcal{S}_{n,m}$ and $\mathcal{S}_{C:n,m}$ are given by
\eqs{
\mathcal{S}_{n,m} \equiv \sum_{\phi} \sum_{\phi^\dag} q_\phi^n  q_{\phi^\dag}^m \, , ~~~~~
\mathcal{S}_{C:n,m} \equiv \sum_{\phi} \sum_{\phi^\dag} \mathcal{C}(\phi, \phi^\dag) q_\phi^n q_{\phi^\dag}^m \, ,
}
where we take the sum over the chiral and anti-chiral superfields independently.
If both $\phi$ and $\phi^\dag$ have color charges, $\mathcal{C}(\phi, \phi^\dag) = (N+1)/2N$, if not, $\mathcal{C}(\phi, \phi^\dag) = 0$.

Finally, the anomalous dimensions for proton decay operators are obtained by substituting the group theoretical values as follows:
\eqs{
[\gamma_{\Ocal^{(1)}}^{(2)}]_{33} & =
[\gamma_{\Ocal^{(2)}}^{(2)}]_{33} = \frac{64}{3} + 8 b_3 \, , \\
[\gamma_{\Ocal^{(1)}}^{(2)}]_{22} & =
[\gamma_{\Ocal^{(2)}}^{(2)}]_{22} = \frac92+3b_2 \, , \\
[\gamma_{\Ocal^{(1)}}^{(2)}]_{YY}
& = \frac{113}{54} + \frac53 b_Y \, , ~~~~
[\gamma_{\Ocal^{(2)}}^{(2)}]_{YY}
= \frac{91}{18} + 3 b_Y \, , \\
[\gamma_{\Ocal^{(1)}}^{(2)}]_{32}
& = 12 \, , ~~~~
[\gamma_{\Ocal^{(2)}}^{(2)}]_{32}
= 20 \, , \\
[\gamma_{\Ocal^{(1)}}^{(2)}]_{2Y}
& = 2 \, , ~~~~
[\gamma_{\Ocal^{(2)}}^{(2)}]_{2Y}
= \frac23 \, , \\
[\gamma_{\Ocal^{(1)}}^{(2)}]_{3Y}
& = \frac{68}{9} \, , ~~~~
[\gamma_{\Ocal^{(2)}}^{(2)}]_{3Y}
= \frac{76}{9} \, .
\label{eq:anomalous_d=6two-loop}
}
This result is consistent with the previous one \cite{Hisano:2013ege}.

\section{Conclusions and Discussion \label{sec:conclusion}}

In this work, we have discussed the two-loop anomalous dimensions for four-Fermi operators in generic supersymmetric models.
Using the two-loop effective K\"ahler potential and the RGEs for the vertex functions with operator insertions, we have derived the generic forms of two-loop anomalous dimensions for the four-Fermi operators.
In particular, we have shown the explicit forms of the anomalous dimensions for $\Del F = 2$ processes and proton decay processes.
The former operators arise from massive vector superfield mediation in $U(1)^\prime$-extended SUSY models, and the latter is predicted in the grand unified theories.
In our study, we consider irreducible decompositions of the effective operators, so that the most of parts of the two-loop anomalous dimensions are written in terms of the group theoretical constants.
While the next-leading order corrections for the proton decay operators have been estimated in some literature \cite{Hisano:2013ege,Hisano:2015ala,Bajc:2016qcc}, the explicit derivation for the two-loop anomalous dimensions has been shown in our study.

Although we have focused on particular processes in this study, it is straightforward to extend our result to $\Del F = 1$ processes.
For the $\Del F = 1$ processes, what we should do is to calculate the fourth line in \cref{eq:2loopano}.
Besides, there remains the operator which would be generated by the massive vector superfield as listed in Ref.~\cite{Piriz:1997id}; $Q^\dag \ovl D^\dag L \ovl E$.
From a phenomenological point of view, this operator would be generated by gauge multiplets with color and weak charges: additional gauge multiplets in the context of grand unified theories.
It is easy to calculate the anomalous dimensions since the chiral part of the operator transforms as a weak doublet under the SM group.
Furthermore, our results would apply to the anomalous dimension for the operators which generate non-holomorphic soft masses since their structure is similar to the one of $\Phi\Psi$ operators.

We have not estimated an effect of them numerically in this work.
Due to the operator mixing via the RGEs, it depends on the initial condition for the Wilson coefficients.
Even though we focus on the $U(1)^\prime$-extended SSMs, the prediction varies with the $U(1)^\prime$ charge assignments.
Since we have also neglected the effects from couplings in superpotential, such as the top-Yukawa coupling, it should be included sooner or later for the sake of completeness.

We note that we have not taken into account mixing with higher-dimensional operators that contain chiral covariant derivatives and/or vector superfields.
Indeed, in general, it is well known that four-Fermi operators mix with dipole-type operators in non-supersymmetric theories.
However, concerning to the operators up to dimension six, such dipole-type operators are written down in K\"ahler potential, e.g. $\int d^4 \theta X^\dag H_u (Q \overleftrightarrow{D_\alpha} \ovl U) \mathcal{W}^{\alpha}$ with a SUSY breaking spurion $X = \theta^2$ \cite{Barbieri:1993av,Elias-Miro:2014eia}.
Therefore, the mixing does not appear in the absence of SUSY breaking.

In order to complete the NLO calculation, we should also take threshold corrections into account since they are the same order corrections.
There would be two threshold scales: one is the initial scale where the massive gauge supermultiplets are integrated out, and another is the mass scale of SUSY particles.
In an appendix of Ref.~\cite{Bajc:2016qcc}, the threshold correction at the SUSY mass scale for supersymmetric four-Fermi operators has been derived in general.
The threshold correction at the initial scale also depends on the UV models, and hence we leave it for future work.

\section*{Acknowledgments}
This work is supported by Grant-in-Aid for Scientific research from the Ministry of Education, Science, Sports, and Culture (MEXT), Japan, No. 16H06492 (for J.H.). The work of J.H. is also supported by World Premier International Research Center Initiative (WPI Initiative), MEXT, Japan.
The work of T.K. is supported by Research Fellowships of the Japan Society for the Promotion of Science (JSPS) for Young Scientists (No.16J04611).

\newpage
\section*{Appendix}
\appendix
\section{Loop Functions \label{eq:loop_func}}
Here, we show the loop functions $\ovl J$ and $\ovl I$.
\eqs{
\ovl J^{c~d~}_{~a~b}(m_1^2, m_2^2) & =\frac{1}{(16\pi)^2} \left[ m_1^2 \left( 1- \ln \frac{m_1^2}{\ovl \mu^2} \right) G^{-1} \right]^c_{~a}
\left[ m_2^2 \left( 1- \ln \frac{m_2^2}{\ovl \mu^2} \right) G^{-1} \right]^d_{~b} \, ,
}
\eqs{
\ovl I^{d~e~f}_{~a~b~c}(m_1^2,m_2^2,m_3^2) &
= \frac12 \frac{1}{(16\pi^2)^2}\left\{ \left( - \frac52 m_1^2 + 4m_1^2 \ln \frac{m_1^2}{\ovl \mu^2} G^{-1} \right)^{d}_{~a} (G^{-1})^e_{~b} (G^{-1})^f_{~c} \right. \\
& - (G^{-1})^d_{~a} \left(m_2^2 \ln\frac{m_2^2}{\ovl \mu^2} G^{-1} \right)^e_{~b} \left(\ln \frac{m_3^2}{\ovl \mu^2} \right)^f_{~c} \\
& - (G^{-1})^d_{~a} \left( \ln\frac{m_2^2}{\ovl \mu^2} G^{-1} \right)^e_{~b} \left(m_3^2\ln \frac{m_3^2}{\ovl \mu^2} \right)^f_{~c} \\
& \left. + (m_1^2 G^{-1})^d_{~a} \left(\ln\frac{m_2^2}{\ovl \mu^2} G^{-1} \right)^e_{~b} \left(\ln \frac{m_3^2}{\ovl \mu^2} \right)^f_{~c}
+\mathrm{cycl.} \right\} \, ,
}
where ``cycl.'' stands for the cyclic permutation of the labels $1,2,3$ and the corresponding indices $a, b, \cdots, f$.
For the loop function $\ovl I$ with adjoint indices $I, J, \cdots$, the inverse of the K\"ahler metric $G^{-1}$ is replaced with $g^2 \del^{IJ}$ where $g$ denotes the corresponding gauge coupling.

\newpage
\bibliography{ref}
\end{document}